\documentclass[aps,prb,showpacs,groupedaddress,twocolumn,amssymb]{revtex4}
\usepackage{bm,amsmath,graphicx}
\begin{document}

\title{Nanomechanical quantum memory for superconducting qubits}

\author{Emily J. Pritchett and Michael R. Geller}

\affiliation{Department of Physics and Astronomy, University of Georgia,  Athens, Georgia 30602-2451}

\date{October 4, 2004}

\begin{abstract}
Many protocols for quantum computation require a quantum memory element to store qubits.  We discuss the accuracy with which quantum states prepared in a Josephson junction qubit can be stored in a nanoelectromechanical resonator and then transfered back to the junction.  We find that the fidelity of the memory operation depends on both the junction-resonator coupling strength and the location of the state on the Bloch sphere.  Although we specifically focus on a large-area, current-biased Josesphson junction phase qubit coupled to the dilatational mode of a piezoelectric nanoelectromechanical disk resonator, many our results will apply to other qubit-oscillator models.  
\end{abstract}


\maketitle
\clearpage

\section{INTRODUCTION}

Nature provides many ways to realize a single qubit, but quantum information processing is useful only if many qubits can be coupled in such a way that two-qubit operations can be performed.  Because qubits must be coherent yet controllable, the macroscopic quantum properties and long coherence times of superconductors make Josephson junctions strong candidates.\cite{MakhlinRMP01} Several proposed quantum computing architectures involve coupling Josephson junction (JJ) flux, phase, or charge qubits together with $LC$ resonators,\cite{MakhlinRMP01,ShnirmanPRL97,MakhlinNat99,MooijSci99,MakhlinJLTP00,YouPRL02,Yukon02,SmirnovPre02,BlaisPRL03,PlastinaPRB03,ZhouPRA04}
superconducting cavities,\cite{Buisson01,BlaisPRA04,PaternostroPRA04,WallraffNat04,GirvinPre03} mechanical resonators,\cite{Cleland&GellerPRL04,Geller&ClelandPre04} or other types of oscillators.\cite{MarquardtPRB01,HekkingPre02,ZhuPRA03,PaternostroPRB04} Such resonator-based coupling schemes have the advantage of additional functionality resulting from the ability to tune the qubits relative to the resonator frequency, as well as to each other. Although harmonic oscillators are ineffective as computational qubits, because the lowest pair of levels cannot be frequency selected by an external driving force, they are quite desirable as coupling elements.

In architectures based on JJs coupled to resonators, the resonators store single qubit states, transfer states from one JJ to another, entangle two or more JJs, and mediate two-qubit quantum logic. In effect, the resonators are the quantum computational analog of the classical memory and bus elements. In this paper we discuss the speed and accuracy with which a state can be stored in a resonator and later retrieved, which depends on both the state being stored and on the coupling strength between the JJ and the resonator.  The model we consider includes no disssipation or decoherence, and any loss of fidelity is a consequence of purely coherent quantum dynamics of the coupled qubit-oscillator system. This issue is essential to consider when designing a real quantum computer.
		
The specific architecture we consider is a large-area, current-biased Josephson junction phase qubit coupled to a nanoelectromechanical resonator.\cite{Cleland&GellerPRL04,Geller&ClelandPre04}  The phase qubit is attractive not only for its robust quantum coherence, but also because an effective method for state preparation, manipulation, and measurement has been developed.\cite{MartinisPRL02}  Our results are obtained by direct numerical integration of the time-dependent Schrodinger equation with a Hamiltionian that is analogous to that of a two-level system in an electromagnetic cavity, and many of our result will apply to other 
qubit-oscillator models. 

\section{SUPERCONDUCTING PHASE QUBIT COUPLED TO NEMS RESONATOR}

The low-energy dynamics of a JJ is determined by the difference $\varphi$ between the phases of the spatially uniform order parameters in the superconductors forming the junction. The Hamiltonian for the system we consider is $H=H_{\rm J} + H_{\rm res} + \delta H$, where $H_{\rm J} \equiv -E_{\rm c} {d^2\over{d\varphi^2}} + U(\varphi)$ is the Hamiltonian of the JJ with current bias $I_{\rm b}$, with $U \equiv -E_{\rm J} (\cos\varphi + s\varphi)$ and $s \equiv I_{\rm b}/I_0$.  $E_{\rm c} \equiv (2e)^2/2C$ is the charging energy and $E_{\rm J} \equiv \hbar I_0/2 e$ is the Josephson coupling energy, with $C$ the junction capacitance and $I_0$ the critical current. In the large-area JJ of interest here, $E_{\rm J}$ is much larger than $E_{\rm c}$. $\omega_{{\rm p}0} \equiv \sqrt{2 E_{\rm c} E_{\rm J}}/\hbar$ is the zero-bias plasma frequency. The lowest two eigenstates, $|0\rangle$ and $|1\rangle$, are used to make a qubit. $H_{\rm res} \equiv \hbar \omega_0a^{\dagger}a$ is the Hamiltonian for the resonator, with $a^\dagger$ and $a$ the creation and annihilation operators for dilatational-mode phonons of frequency $\omega_0$. The resonator is a piezoelectric disk sandwiched between two capacitor plates. Finally, the interaction term is ${\delta}H=-ig(a - a^{\dagger}) \varphi,$ where $g$ is a coupling constant with dimensions of energy that depends on the geometric and material properties of the resonator.\cite{Cleland&GellerPRL04,Geller&ClelandPre04} This Hamiltonian, which is analogous to a few-level atom in a single-mode electromagnetic cavity, is discussed further in Refs.~[\onlinecite{Cleland&GellerPRL04}] and [\onlinecite{Geller&ClelandPre04}].

The junction Hamiltonian $H_{\rm J}$ depends on the dimensionless bias current $s$, which is time-dependent. We expand the state of the coupled system in a basis of instantaneous eigenstates $|mn\rangle_{\! s}$ of $H_0 \equiv H_{\rm J} + H_{\rm res}$, defined by $H_0(s) |mn\rangle_{\! s} = E_{mn}(s) |mn\rangle_{\! s},$ where $|mn\rangle_{\! s} \equiv |m\rangle_{\rm J} \otimes |n\rangle_{\rm res},$ with $|m\rangle_{\rm J}$ and $|n\rangle_{\rm res}$ the eigenstates of the uncoupled JJ and resonator, respectively. The wave function is then expanded as
\begin{equation}
|\psi(t)\rangle = \sum_{mn} c_{mn}(t) \, e^{-(i/\hbar) \! \int_{t_0}^t dt' E_{mn}(s)} |mn\rangle_{\! s}.
\label{general wave function expansion}
\end{equation}
The probability amplitudes in the instantaneous interaction representation satisfy
\begin{eqnarray}
i \hbar {\dot c}_{mn} &=& \sum_{m'n'} \langle mn | \delta H - i \hbar \partial_t |m'n'\rangle_{\! s}  \nonumber \\
&\times& e^{(i/\hbar) \! \int_{t_0}^t \! dt' [E_{mn}(s) - E_{m'n'}(s)]} \ c_{m'n'}.
\label{schrodinger equation}
\end{eqnarray}
All effects of dissipation and decoherence are assumed to be negligible over the time scales studied here.

We will assume that the JJ states are well approximated by harmonic oscillator eigenfunctions, which is an excellent approximation unless $s$ is very close to unity. 
Transitions between the instantaneous eigenstates are caused by nonadiabatic variation of $s$, through the term $ \langle mn | {\textstyle{\partial \over \partial t}} |m'n'\rangle_{\! s} = \langle mn | {\textstyle{\partial \over \partial s}} |m'n'\rangle_{\! s} \, {\dot s},$ which can be evaluated analytically in the harmonic-oscillator limit.\cite{Geller&ClelandPre04}

\section{NEMS RESONATOR AS A QUANTUM MEMORY ELEMENT}

An arbitrary qubit state 
\begin{equation}
|\psi\rangle_{\rm J} = \alpha |0\rangle_{\rm J} + \beta |1\rangle_{\rm J}
\label{qubit state}
\end{equation}
prepared in the JJ can be stored in the ground and one-phonon states of the resonator's dilatational mode as follows: Assuming the resonator is initially in the ground state and the JJ is detuned from the resonator, the coupled system is prepared in the initial state
\begin{equation}
\big(\alpha |0\rangle_{\rm J} + \beta |1\rangle_{\rm J} \big) \! \otimes \! |0\rangle_{\rm res} = \alpha |00\rangle + \beta |10\rangle,
\label{storage initial state}
\end{equation}
with $|\alpha|^2 + |\beta|^2 =1.$ The bias current $s$ is then adiabatically varied to tune the JJ level spacing $\Delta \epsilon = \hbar \omega_{{\rm p}0} (1-s^2)^{1/4}$ to $\hbar \omega_0$, reaching the resonant value $s^* \equiv \sqrt{1-(\omega_0/\omega_{{\rm p}0})^4}$ at time $t=0$. Neglecting any nonadiabatic corrections, the probability amplitudes in the instantaneous interaction representation at this time are 
\begin{equation}
c_{mn}(0) = \big( \alpha \, \delta_{m0} + \beta \, \delta_{m1} \big) \delta_{n0}. 
\label{storage initial condition}
\end{equation}

If the interaction strength $g$ is small compared with the level spacing $\Delta \epsilon$, the subsequent dynamics is well described in the rotating-wave approximation (RWA) of quantum optics.\cite{Scully} In this approximation, and on resonance, we can write Eq.~(\ref{schrodinger equation}) as
\begin{eqnarray}
{\dot c}_{0n} &=& \frac{g}{\hbar} \, \sqrt{n} \  x_{01} \, c_{1,n-1} \nonumber \\
{\dot c}_{1n} &=&  -\frac{g}{\hbar} \, \sqrt{n+1} \  x_{01} \, c_{0,n+1},
\label{RWA equations}
\end{eqnarray}
where $x_{01} \equiv \langle 0 | \varphi | 1\rangle_{\rm J} = \ell^* /\sqrt{2}$ is an effective dipole moment. Here $\ell^* \equiv (2 E_{\rm c} / E_{\rm J})^{\! {1\over4}} \! (1 - {s^*}^2)^{-{1\over8}}$ is the characteristic width in $\varphi$ of the harmonic oscillator eigenfunctions in the JJ, when tuned to the resonator.  Using Eqs.~(\ref{storage initial condition}) and (\ref{RWA equations}) we obtain, for $t \ge 0,$
\begin{eqnarray}
c_{00}(t) &=& \alpha \nonumber \\
c_{01}(t) &=& \beta \, \sin({\textstyle{\Omega t \over 2}})  \nonumber \\
c_{10}(t) &=& \beta \, \cos({\textstyle{\Omega t \over 2}}) \nonumber \\
c_{11}(t) &=& 0,
\end{eqnarray}
and all $c_{mn}(t)$ with $n>1$ equal to zero. $\Omega \equiv 2 g x_{01} / \hbar$ is the resonant vacuum Rabi frequency. 

After a time $\Delta t = \pi / \Omega$, the nonvanishing probability amplitudes are
\begin{eqnarray}
c_{00}(t) &=& \alpha \nonumber \\
c_{01}(t) &=& \beta , 
\label{storage amplitudes}
\end{eqnarray}
corresponding to the interaction-representation state $|0\rangle_{\rm J} \otimes (\alpha |0\rangle_{\rm res} + \beta |1\rangle_{\rm res}).$ In this sense, the qubit state of Eq.~(\ref{qubit state}) has been stored the in the resonator's vacuum and one-phonon states $|0\rangle_{\rm res}$ and $|1\rangle_{\rm res}$. The JJ is now adiabatically detuned from the resonator.

To retrieve the stored state, we again bring the systems into resonance at time $t_1$. Using the stored amplitudes of Eq.~(\ref{storage amplitudes}) as initial conditions, and assuming that $t \ge t_1$, the RWA equations now lead to
\begin{eqnarray}
c_{00}(t) &=& \alpha \nonumber \\
c_{01}(t) &=& \beta \, \cos[ {\textstyle{\Omega \over 2}}(t-t_1)]  \nonumber \\
c_{10}(t) &=& - \beta \, \sin[{\textstyle{\Omega \over 2}}(t-t_1)],
\end{eqnarray}
the others vanishing. This time the systems are held in resonance for an interval $\Delta t = 3 \pi /\Omega$, after which the original state (\ref{storage initial state}) is recovered.

The above analysis, which is based on the adiabatic approximation and the RWA, suggests that perfect memory performance can be obtained with arbitrarily fast gate times. This is incorrect, of course, because the actual quantum memory performance is controlled by the corrections to these approximations, which we shall analyze numerically in the following section.

\section{QUANTUM MEMORY FIDELITY}

The accuracy of a storage and retrieval operation can be characterized by the absolute value of the overlap between the intended and achieved final states, or the fidelity $F$ of the memory operation. Accounting for the fact that the intended and actual final state have the same phase factors resulting from the time-evolution of the instantaneous eigenstates, $F$ is given by the absolute value of the inner product of the intended and achieved interaction-representation probability amplitudes. In our results we actually plot the fidelity squared, 
\begin{equation}
F^2 = \big| \alpha^* c_{00}(t_{\rm f}) + \beta^* c_{10}(t_{\rm f}) \big|^2 \! ,
\label{F definition}
\end{equation}
which characterizes the {\it probability} that the memory device operates correctly. The fidelity will have a weak dependence on $t_{\rm f}$, which we shall discuss below.

\begin{figure}
\includegraphics[width=9.0cm]{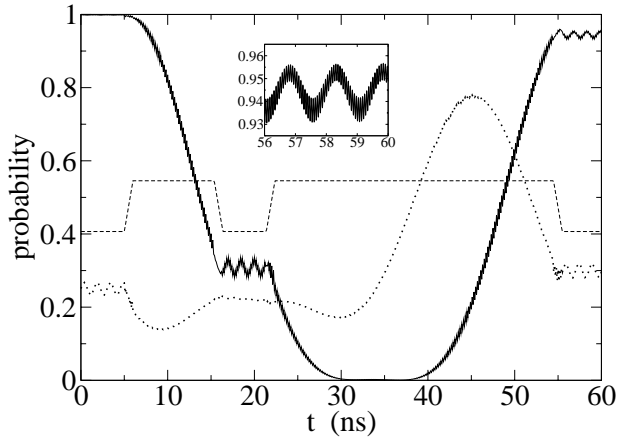}
\caption{Storage and retrieval of the state $2^{-{1 \over 2}}(|0\rangle+|1\rangle).$ The solid curve is the overlap squared with the inital state. After $60 \, {\rm ns}$ the qubit is successfully retrieved with a squared fidelity of about 94\%. The dotted curve gives the occupation of the state $2^{-{1 \over 2}}(|00\rangle+|01\rangle)$ in which the qubit is stored in the resonator. The dimensionless interaction strength $g/\hbar \omega_0$ here is $5\%.$ The dashed curve is the bias current $s$. (inset) Expanded view of the weak  $t_{\rm f}$-dependence of the fidelity.}
\label{memory figure}
\end{figure}

In Fig.~\ref{memory figure} we show the results of simulating the storage and retrieval of the qubit state $2^{-{1\over2}}(|0\rangle+|1\rangle),$ which is on the equator of the Bloch sphere. The JJ is that of Ref.~[\onlinecite{MartinisPRL02}], with parameters $E_{\rm J} = 43.05 \, {\rm meV}$ and $E_{\rm c} = 53.33 \, {\rm neV}$, and the resonator has a dilatational-mode frequency $\omega_0/2 \pi$ of  $15 \, {\rm GHz}.$ $g$ is $0.05 \, \hbar \omega_0.$ The dimensionless bias current on resonance is $s^* = 0.545,$ and the off-resonant value of $0.407$ was determined by optimizing the fidelity. The memory operation of Fig.~\ref{memory figure} was achieved with a squared fidelity of about 94\%.

The inset to  Fig.~\ref{memory figure} shows the final stages of the occupation of the state $2^{-{1 \over 2}}(|00\rangle+|10\rangle).$ Unless the state being considered happens to be an eigenstate of the coupled JJ-resonator system, the fidelity will depend on $t_{\rm f}.$ However, the variation is quite small, so it is reasonable to assign an average fidelity, which is what we do below.
 
\subsection{Memory fidelity versus coupling strength}

\begin{figure}
\includegraphics[width=8.0cm]{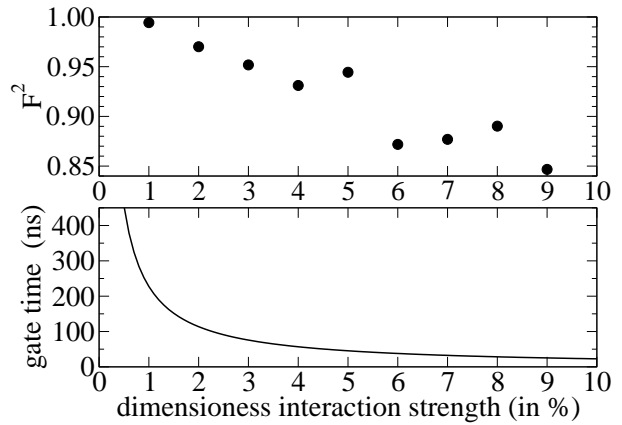}
\caption{(upper panel) Memory fidelity for equator state $2^{-{1\over2}}(|0\rangle+|1\rangle)$ as a function of $g/\hbar \omega_0$, in percent. (lower panel) Time needed to store and retrieve state as a function of $g/\hbar \omega_0.$}
\label{Fvsg figure}
\end{figure}

The memory operation shown in Fig.~\ref{memory figure} takes two vacuum Rabi periods---about $45 \, {\rm ns}$---to complete, disregarding times during which the system is detuned. The gate speed can be increased by increasing $g$, but then the fidelity decreases. In the upper panel of Fig.~\ref{Fvsg figure} we show the memory fidelity squared for the qubit state $2^{-{1 \over 2}}(|0\rangle+|1\rangle)$ as a function of $g/\hbar \omega_0.$ As expected, the fidelity gradually decreases with increasing $g.$ However, there are deviations from strictly monotonic dependence on $g,$ which we believe to be consequences of our use of a fixed (not optimized for each $g$) nonresonant value of $s$. 
The lower panel of Fig.~\ref{Fvsg figure} gives the gate time as a function of $g/\hbar \omega_0,$ again disregarding nonresonant evolution. These results suggest that memory fidelities better than 90\% can be achieved using phase qubits and resonators with coherence times of a few hundred ns.

\subsection{State dependence of memory fidelity}

\begin{figure}
\includegraphics[width=8.0cm]{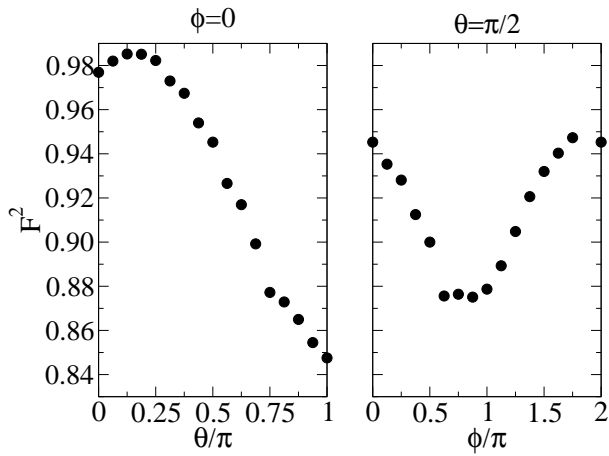}
\caption{State dependence of memory fidelity, for the same JJ-resonator system studied in Fig.~\ref{memory figure}, with $g/\hbar \omega_0 = 0.05.$ (left) Fidelity squared as a function of $\theta,$ along the arc $\phi=0$ on the Bloch sphere. (right) Fidelity squared as a function of $\phi$ around the equator.}
\label{Bloch figure}
\end{figure}

Interestingly, the memory fidelity depends not only on the strength of the JJ-resonator interaction, but also on the qubit state itself. This is because, as we explained above, the fidelity is determined by the corrections to the adiabatic and rotating-wave approximations, and these corrections are state dependent. Using a Bloch sphere representation 
\begin{equation}
|\psi\rangle = \cos({\textstyle{\theta \over 2}}) \, |0\rangle + \sin({\textstyle{\theta \over 2}}) e^{i\phi} \, |1\rangle
\label{Bloch qubit state}
\end{equation}
for the stored qubit, we show in Fig.~\ref{Bloch figure} the memory fidelity along two great circles, from $|0\rangle$ to $|1\rangle$ along $\phi=0$ (left) and around the equator (right) starting and finishing at $2^{-{1\over2}}(|0\rangle+|1\rangle).$ 

The dependence of $F^2$ on $\theta$ can be understood as follows: When $\theta=0$, the initial state of the coupled system is $|00\rangle$, because the resonator always starts in the ground state. For a weakly coupled system, $|00\rangle$ is close to the exact ground state for any $s$, because there are no other $|mn\rangle$ states degenerate with $|00\rangle$. In the adiabatic $ds/dt \rightarrow 0$ limit, the large component of $|00\rangle$ in the exact instantaneous ground state will remain there with unit probability, a consequence of the adiabatic theorem, leading to a high memory fidelity for the qubit state $|0\rangle$. The $|1\rangle$ state, by contrast, derives no protection from the adiabatic theorem and is subject to errors caused by the corrections to the RWA. The weaker $\phi$ dependence, which favors equator states pointing in the $2^{-{1\over2}}(|0\rangle+|1\rangle)$ direction, or in a spin-${1 \over 2}$ language with  $|0\rangle=| \! \! \uparrow\rangle$ and $|1\rangle=| \! \! \downarrow\rangle,$ equator states pointing in the $x$ direction, is not understood at present.

\section{DISCUSSION}

We have explored the accuracy with which piezoelectric nanoelectromechanical resonators can be used to store qubit states prepared in a current-biased Josephson junction.    We find that the memory fidelity depends on both the resonator-JJ interaction strength as well as the position of the state on the Bloch sphere. Finding the optimum balance between fidelity and gate operation time will be essential in the design large-scale superconducting quantum computers. Overall, our simulations suggest that after further optimization, generic states on the Bloch sphere should be able to be stored and retrieved in a few hundred ns with accuracies better than $90\%,$ which would be a significant accomplishment in experimental quantum information processing. 

We expect that many of the results presented here will apply to other qubit-oscillator systems as well. In particular, the memory operations discussed here could be carried out with the superconducting transmission line resonator architecture currently being developed at Yale.\cite{BlaisPRA04,WallraffNat04,GirvinPre03} 

\acknowledgments

This work was supported by the National Science Foundation under CAREER Grant No.~DMR-0093217. We thank Andrew Cleland, Steve Lewis, and Andrew Sornborger for useful discussions.

\bibliography{/Users/mgeller/Papers/bibliographies/MRGqc,/Users/mgeller/Papers/bibliographies/MRGpre,/Users/mgeller/Papers/bibliographies/MRGbooks,/Users/mgeller/Papers/bibliographies/MRGgroup,/Users/mgeller/Papers/bibliographies/MRGnano}

\end{document}